\documentclass[pra,aps,twocolumn,groupedaddress,superscriptaddress,floatfix,showpacs]{revtex4}
\usepackage{epsfig}
\usepackage{amsmath}
\usepackage{mathrsfs}
\usepackage{textcmds}
\allowdisplaybreaks[1]
\newcommand{\be}{\begin{equation}}
\newcommand{\ee}{\end{equation}}
\newcommand{\bea}{\begin{eqnarray}}
\newcommand{\eea}{\end{eqnarray}}
\newcommand{\ket}[1]{\left|#1\right\rangle}
\newcommand{\bra}[1]{\left\langle #1\right|}

\newcommand{\bc}{\begin{center}}
\newcommand{\ec}{\end{center}}

\renewcommand{\(}{\left(}
\renewcommand{\)}{\right)}
\renewcommand{\[}{\left[}
\renewcommand{\]}{\right]}
\newcommand{\forget}[1]{}
\newcommand{\re}{{\rm e}}
\newcommand{\ri}{{\rm i}}
\newcommand{\half}{\dfrac{1}{2}}
\begin{document}
\title{Subwavelength atom localization via amplitude and phase control of
the absorption spectrum-II}
\author{Kishore T. Kapale}
\email{Kishor.T.Kapale@jpl.nasa.gov}
\affiliation{Jet Propulsion Laboratory,
California Institute of Technology, Mail Stop 126-347, 4800 Oak Grove Drive, 
Pasadena, California 91109-8099}

\author{M. Suhail Zubairy}
\email{zubairy@physics.tamu.edu}
\affiliation{Institute for Quantum Studies and Department of
Physics, Texas A\&M University, College Station, TX 77843-4242}

\begin{abstract}
Interaction of the internal states of an atom with spatially dependent standing-wave cavity field can impart position information of the atom passing through it leading to subwavelength  atom localization. We recently demonstrated a new regime of atom localization [Sahrai {\it et al.}, Phys. Rev. A {\bf 72}, 013820 (2005)], namely sub-half-wavelength localization through phase control of electromagnetically induced transparency. This regime corresponds to extreme localization of atoms within a chosen 
half-wavelength region of the standing-wave cavity field. Here we present further investigation of the simplified model considered earlier and show interesting features of the proposal. We show how the model can be used to simulate variety of energy level schemes. Furthermore, the dressed-state analysis is employed  to explain the emergence and suppression of the localization peaks, and the peak positions and widths.
The range of parameters for obtaining clean sub-half-wavelength localization is identified.
\end{abstract}
%

\maketitle
\section{Introduction}
Precise localization of atoms has attracted considerable attention in recent years. Optical manipulations allow probing the center-of-mass degrees of freedom of atoms with subwavelength precision. The interest in subwavelength atom localization 
is largely due to its applications to many areas requiring manipulations of atomic center-of-mass  degrees of freedom,  such as laser cooling~\cite{ChuMetcalf},
Bose-Einstein condensation~\cite{Collins96}, and atom
lithography~\cite{lithography} alongwith fundamentally important issues such as  measurement of the center-of-mass wavefunction of moving atoms~\cite{KapaleWavefunction}.   

Optical techniques for position measurements of the atom are of considerable interest from both theoretical and experimental point of view.  Several scheme have been proposed for the localization of an atom using optical methods~\cite{welch9193}. It is well known that optical methods provide better spatial resolution in position measurement of the atom.  For example, in the optical virtual slits scheme the atom interacts
with a standing-wave field and  imparts a phase shift to the field. Measurement of this phase shift then gives the position information of the atom~\cite{WallsZollerWilkens}.
In another related idea based on phase quadrature measurement is
considered in Ref.~\cite{Walls95}. Kunze {\it et al.}~\cite{Kunze97} demonstrated how 
the entanglement between the atomic position and its internal state
allows one to localize the atom without directly affecting its spatial 
wave function. It is shown that, by using Ramsey
interferometry, the use of a coherent-state cavity field is better
than the classical field to get a higher resolution in position
information of the atom~\cite{Kien97}.  Resonance imaging methods have also been employed in experimental studies of the precision position measurement of the moving atoms~\cite{Thomas90,Bigelow97}.

More recently, atom-localization
methods based on the detection of the spontaneously emitted photon
during the interaction of an atom with the classical standing-wave
field, are considered~\cite{Zoller96,Herkomer97,QamarPRA2000,QamarOC2000}. 
It is, however,  important to note that from an 
experimental point of view, observation of spontaneous emission spectrum is very
tricky and difficult. In this context, another scheme based on a three-level
$\Lambda$-type system interacting with two fields, a probe
laser field and a classical standing wave coupling field, is used
for atom localization by Paspalakis and Knight~\cite{Knight2001}. 
They observe that in the case of a weak  probe field, measurement of the population in the upper level leads to sub-wavelength localization of the atom during its motion in the standing wave.  Thus, in essence, this scheme uses absorption of a probe field for atom localization.  Atomic coherence effects, such as coherent population trapping, have also been shown to be useful for subwavelength localization of atoms by Agarwal and Kapale~\cite{AgarwalKapaleAL}, where monitoring the coherence of the trapping state gives rise to subwavelength localization of an atom to a precision prechosen through the ratio of the square of Rabi frequencies of the strong standing-wave drive field and a weak probe field. 

The authors (with collaborators) recently proposed a subwavelength atom localization scheme through phase control of the absorption of a weak probe field by the atom.  A modified $\Lambda$-type level scheme with an extra level and the drive fields forming a complete loop was shown to  introduce a phase dependence in the response of the atomic medium to a weak probe field.  This phase controllable atomic response was shown to give rise to tunable group velocity from subluminal to superluminal in a single system~\cite{Sahrai:2004}.  By considering one of the drive fields to be a standing-wave field of a cavity it was shown that the same scheme can be used to localize an atom flying through the standing wave field to subwavelength domain~\cite{Sahrai:2005}. 

This article is a sequel to the earlier article~\cite{Sahrai:2005}, henceforth referred to as $\mathscr{I}$. In $\mathscr{I}$  a restricted parameter range of the model was considered  to show the possibility of sub-half-wavelength localization. In this article  further investigations of the analytical results are carried out to show how the model can be used to simulate variety of atomic systems with varying energy level spacings, different atomic dipole matrix elements and decay properties. Appropriate parameters required to obtain different regimes of localization are studied in detail. A dressed-states approach is also considered to give insight into the results obtained.

The article is organized as follows: For completeness, a brief description of the procedure to determine the susceptibility of the atom to a weak probe field is given in Sec.~\ref{Sec:Model}. Then,  in Sec.~\ref{Sec:Results} A, the susceptibility expression is studied in detail to arrive at the conditions for observing atom localization. Various parameter ranges for the drive field Rabi frequencies and decay properties are considered in order to simulate variety of atomic species and to clarify experimentally controllable features and properties of the model and numerical results and their explanation through the analytical probe susceptibility expression is presented in Sec.~\ref{Sec:Results} B. A simple dressed states treatment is presented in Sec.~\ref{Sec:Results}C  in order to explain the results obtained in the earlier sections. Finally the conclusion is presented.

\section{The Model and Equations}
\label{Sec:Model}
\begin{figure}[ht]
\centerline{\includegraphics[width=\columnwidth]{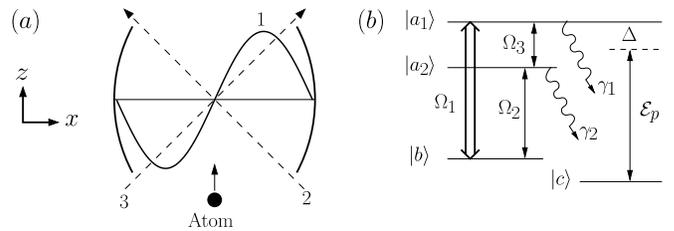}}
\caption{\label{Fig:Scheme}The Model: $(a)$ The cavity supports the standing
wave field (1) corresponding to Rabi frequency $\Omega_1$. Two other fields
(2, 3) are applied at an angle as shown. The atom enters the cavity along the $z$
axis and interacts with the three drive fields. 
The whole process takes place in the $x$-$z$ plane. $(b)$ The energy level structure
of the atom. Probe field, denoted by $\mathcal{E}_p$, is detuned by an amount
$\Delta$ from the $\ket{a_1}-\ket{c}$ transition. The fields (2, 3) shown in $(a)$
part of the figure correspond to the fields with Rabi frequencies $\Omega_2$ and
$\Omega_3$ respectively. The decay rates from the upper levels $\ket{a_1}$ and
$\ket{a_2}$ are taken to be $\gamma_1$ and $\gamma_2$ respectively.}
\end{figure}

The schematics of the proposed scheme are shown in Fig.~\ref{Fig:Scheme}. We
consider an atom, moving in the $z$ direction, as it passes through a
classical standing-wave field of a cavity. The cavity is taken to be aligned 
along the $x$ axis. The internal energy level structure of the atom is shown in 
Fig.~\ref{Fig:Scheme}(b). 
The radiative decay rates from the level $\ket{a_1}$ and
 $\ket{a_2}$ to level $\ket{c}$ are taken to be $\gamma_1$ and $\gamma_2$.
The upper level $\ket{a_1}$ is coupled to the level
$\ket{a_2}$ and further the level $\ket{a_2}$ is coupled to level
$\ket{b}$ via classical fields with
Rabi frequencies $\Omega_3$ and $\Omega_2$, respectively. 
In addition, the upper level $\ket{a_1}$ is 
coupled to level
$\ket{b}$ via a classical standing-wave field having Rabi frequency $\Omega_1$. It should be noted that the Rabi frequency of the standing wave is  position dependent and is taken to be $\Omega_1(x)=\Omega_1 \sin \kappa x$ . Here $\Omega_1(x)$ is defined to include the  position 
dependence and $\kappa$ is the wave vector of the standing wave
field, defined as $\kappa =2 \pi/{\lambda}$, where $\lambda$ is the wavelength of the standing-wave field of the cavity.
We assume that the atom is initially in the state $\ket{c}$  and
interacts with a weak probe field that is near resonant with
$\ket{c}\rightarrow\ket{a_1}$ transition. The detuning of the probe field  on this transition is taken to be $\Delta$. We  assume that the
center-of-mass position distribution of the atom is nearly uniform along the
direction of the standing wave. Therefore, we apply the 
Raman-Nath
approximation and neglect the kinetic part of the atom from the
Hamiltonian~\cite{Meystre:1999}. Under these circumstances, the Hamiltonian 
of the system in the
rotating wave approximation can be written as
\begin{equation}
\label{eq:hamiltonian1} \mathcal{H} = \mathcal{H}_0 +
\mathcal{H}_I
\end{equation}
where
\begin{equation}
\label{eq:hamiltonian2}\mathcal{H}_0 = \hbar
\omega_{a_1}\ket{a_1}\bra{a_1} + \hbar
\omega_{a_2}\ket{a_2}\bra{a_2} + \hbar\omega_b\ket{b}\bra{b} +
\hbar\omega_c\ket{c}\bra{c},
\end{equation}
and
\begin{multline}
\label{eq:hamiltonian3}\mathcal{H}_I=-\frac{\hbar}{2} \left[
\Omega_1 \re^{- \ri \nu_1 t} \sin{\kappa x}\, \ket{a_1}\bra{b}\right.
\\
+ \Omega_2 \re^{\ri k x \cos \theta_2} \re^{- \ri
\nu_2 t}\ket{a_2}\bra{b}  \\ 
+ \left.\Omega_3 \re^{\ri k x \cos{\theta_3}}  \re^{- \ri \nu_3
t}\ket{a_1}\bra{a_2}
+\frac{\mathcal{E}_p\wp_{a_1 c}}{\hbar} \re^{- \ri
\nu_p t}\ket{a_1}\bra{c}\right]+ \mbox{H.c.}
\end{multline}
Here $\omega_i$ are the frequencies of the states
$\ket{i}$ and $\nu_i$ are the frequencies of the 
optical fields, and  $\theta_{2}$, $\theta_{3}$ are the angles made by the propagation direction of the fields $\Omega_{2}$ and $\Omega_{3}$ with respect the $x$ axis respectively. The
subscript $p$ stands for the quantities corresponding to the probe
field\mdash i.e., ${\mathcal{E}_p}$ and $\nu_p$ are the amplitude and 
frequency of the probe field. Also $\wp_{a_1 c}$ is the 
dipole matrix element of the $\ket{c}\rightarrow
\ket{a_1}$ transition. For simplicity, we assume that
the Rabi frequencies $\Omega_1$ and $\Omega_2$ are 
real and $\Omega_3$ is complex\mdash i.e., $\Omega_3=\Omega_3 \re^{- \ri \varphi }$. This choice of imparting a carrying phase to field 3, is only for the convenience of calculations.  As will become clear later, only the relative phase of the three fields is 
important and absolute phases do not matter.
The dynamics of the system is
 described using density matrix approach as:
\begin{equation}
\dot{\rho} = - \frac{\ri}{\hbar} [H,\rho]  - \frac{1}{2}
\{\Gamma,\rho\},
\end{equation}
where $\{\Gamma,\rho\} = \Gamma\rho + \rho \Gamma$. Here the decay
rate is incorporated into the equation by a relaxation matrix
$\Gamma$, which is defined, by the equation $\langle n| \Gamma |
m \rangle = \gamma_n \delta_{nm}$. The detailed calculations of
these equations are given in the Appendix of $\mathscr{I}$. 

Our goal is to obtain information about the atomic position from the susceptibility of 
the system at the probe frequency.  Therefore, we need to determine the steady state value of the off-diagonal the density matrix element,  $\rho_{a_1 c}$. After necessary algebraic calculation and moving to appropriate rotating frames, we obtain a set of 
density
matrix equations.  To determine $\rho_{a_1 c}$  we only need following equations
\begin{align}
\dot{\tilde{\rho}}_{a_1 c} &= - [\ri (\omega_{a_1 c}-\nu_p) +
\half\gamma_1]\tilde{\rho}_{a_1 c} + \frac{\ri}{2} \Omega_3 \re^{- \ri \varphi }  \re^{ \ri k x
\cos \theta_3 } \tilde{\rho}_{a_2 c} \nonumber \\ 
&\qquad+  \frac{\ri}{2} \Omega_1
\sin \kappa x \tilde{\rho}_{b c} - \ri
\frac{\mathcal{E}_p \wp_{a_1 c} }{2 \hbar} (\tilde{\rho}_{a_1 a_1}
- \tilde{\rho}_{cc}), 
\nonumber \\ 
\dot{\tilde{\rho}}_{a_2 c} &= -[\ri (\omega_{a_2 c} - (\nu_p- \nu_3)) + \half\gamma_2] \tilde{\rho}_{a_2 c} 
\nonumber \\
&\qquad+ 
  \frac{\ri}{2} \Omega_2 \re^{ \ri k x \cos
\theta_2 }\tilde{\rho}_{b c}  +
\frac{\ri}{2} \Omega_3 \re^{ \ri \varphi } \re^{- \ri k x \cos \theta_3 }
\tilde{\rho}_{a_1 c}
\nonumber \\ 
&\qquad \qquad- \ri \frac{\mathcal{E}_p \wp_{a_1 c}}{2 \hbar} \tilde{\rho}_{a_2 a_1}, 
\nonumber \\
\dot{\tilde{\rho}}_{b c} &= - [\ri ( \omega_{b c} + \nu_1 -\nu_p ) +
\gamma_{b c}]\tilde{\rho}_{b c} + \frac{\ri}{2} \Omega_1 \sin \kappa x \tilde{\rho}_{a_1 c} 
\nonumber \\
&\qquad+ \frac{\ri}{2}\Omega_2 \re^{- \ri k x \cos \theta_2 } \tilde{\rho}_{a_2 c} - \ri
\frac{\mathcal{E}_p \wp_{a_1 c}}{2\hbar} \tilde{\rho}_{b a_1}.
\label{eq:tilderhodot}
\end{align}

\begin{figure*}[ht]
\centerline{\includegraphics[width=\linewidth]{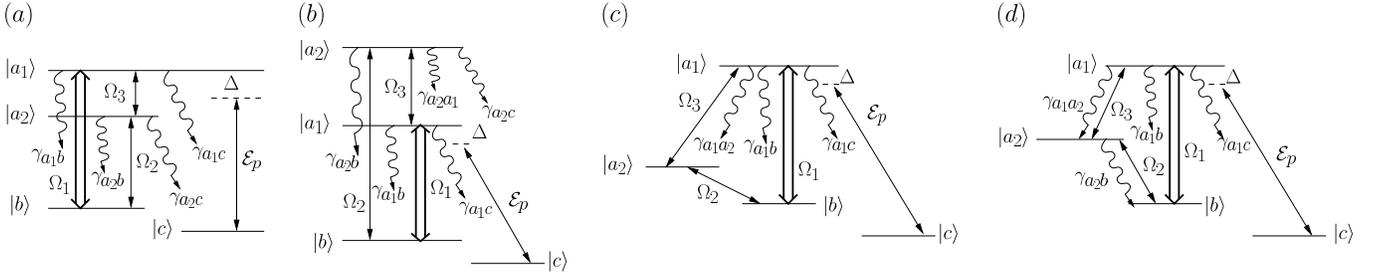}}
\caption{\label{Fig:levelschemes} Several level schemes that can be studied using our model in Fig. 1$(b)$, within the weak probe limit. The decay rates are defined as follows: $(a)$ $\gamma_1 = \gamma_{a_1b} + \gamma_{a_1c}$ and $\gamma_2=\gamma_{a_2b}+\gamma_{a_2c}$; $(b)$ $\gamma_1 = \gamma_{a_1b} + \gamma_{a_1c}$ and $\gamma_2 = \gamma_{a_2a_1}+\gamma_{a_2b} + \gamma_{a_2c}$; $(c)$ $\gamma_1 =\gamma_{a_1a_2}+ \gamma_{a_1b} + \gamma_{a_1c}$ and $\gamma_2=0$;  $(d)$ $\gamma_1 =\gamma_{a_1a_2}+ \gamma_{a_1b} + \gamma_{a_1c}$ and $\gamma_2=\gamma_{a_2b}$. Here $\gamma_{ij}$ corresponds to spontaneous decay rate from level $\ket{i}$ to level $\ket{j}$. It can be noted that the positions of levels $\ket{a_1}$ and $\ket{a_2}$ in $(b)$ are reversed compared to the other schemes. Slight modifications in the equations are needed to simulate level scheme in $(b)$ with the equations given in the text. The transformations required is:  the complex Rabi frequency $\Omega_3 \rightarrow \Omega_{3}^{*}$---i.e., $\nu_3 \rightarrow - \nu_3$. It can be easily shown that the results remain unchanged under these transformations, as discussed in the text.}
\end{figure*}

As we know, the dispersion and absorption are related to the
susceptibility of the system and is determined by $\rho_{a_1 c}$.
We take the probe field to be weak, and  calculate the
polarization of the system to lowest order in $\mathcal{E}_p$.
We keep all the terms of the driving fields but keep
only linear terms in the probe field. The atom is 
initially in the ground state $\ket{c}$, therefore we use
\begin{equation}
\tilde{\rho}_{cc}^{(0)} =1, \quad \tilde\rho_{b a_1}^{(0)}=0,
\quad \tilde\rho_{a_2 a_1 }^{(0)} = 0, \quad \tilde\rho_{a_1 a_1}^{(0)} = 0.
\label{Eq:initial}
\end{equation}
Equation~(\ref{eq:tilderhodot}) can then be simplified considerably to obtain
\begin{align}
\dot{\tilde{\rho}}_{a_1 c} &= - (\ri \Delta +
\half\gamma_1)\tilde{\rho}_{a_1 c} + \frac{\ri}{2} \Omega_3\, \re^{- \ri \varphi } \re^{ \ri k x
\cos \theta_3 } \tilde{\rho}_{a_2 c} 
\nonumber \\
&\qquad+  \frac{\ri}{2} \Omega_1 \sin{\kappa x} \,\tilde{\rho}_{b c} + \ri
\frac{\mathcal{E}_p \wp_{a_1 c} }{2 \hbar} , 
\nonumber \\
\dot{\tilde{\rho}}_{a_2 c} &= - (\ri \Delta + \half\gamma_2)\tilde{\rho}_{a_2 c} 
+ \frac{\ri}{2} \Omega_3\, \re^{\ri \varphi } \re^{- \ri k x \cos \theta_3 } \tilde{\rho}_{a_1 c} 
\nonumber \\
&\qquad+  \frac{\ri}{2} \Omega_2 \re^{\ri k x \cos \theta_2 }\tilde{\rho}_{b c}, \nonumber \\
\dot{\tilde{\rho}}_{b c} &= - \ri\, \Delta\, \tilde{\rho}_{b c} +
\frac{\ri}{2} \Omega_1  \sin{\kappa x} \,\tilde{\rho}_{a_1 c}
\nonumber \\
&\qquad +  \frac{\ri}{2} \Omega_2 \re^{- \ri k x \cos
\theta_2 } \tilde{\rho}_{a_2 c}. \label{eq:tilderhodot2}
\end{align}
Here we have introduced the detuning of the probe field and the 
frequency difference between levels $\ket{a_1}$ and $\ket{c}$,
\begin{equation}
\Delta = \omega_{a_1c} - \nu_p = \omega_{a_2c}+ \nu_3 - \nu_p =
\omega_{bc} + \nu_1 - \nu_p.
\label{Eq:Delta}
\end{equation}
Here we have also assumed that $\gamma_{b c}=0$.
It can be easily seen that these set of equations can also be used to simulate a variety of level schemes as shown in Fig.~\ref{Fig:levelschemes}, after redefining the decay rates accordingly as discussed in the caption. The schme in Fig.~\ref{Fig:levelschemes}$(b)$ requires special attention as the positions of the states $\ket{a_1}$ and $\ket{a_2}$ are reversed in the order of increasing energy compared to the other levelschemes. This entails small change in the rotating frame that is chosen to arrive at the simplified density matrix equations. The transformation required can be accomplished by replacing the complex Rabi frequency $\Omega_3$ by its complex conjugate $\Omega_3^*$ and changing its frequency $\nu_3$ to $-\nu_3$. The density matrix equations so obtained are identical to the set~\eqref{eq:tilderhodot2} given above except for the redifinition of the phase from $\varphi\rightarrow -\varphi$. However, as will be seen later, the phase enters through the term $\cos\varphi$ in the response of the atoms to a weak probe field, thus the final results  are identical for all the models discussed in Fig.~\ref{Fig:levelschemes}

This set of equations can be solved analytically; the detailed discussion can be found in the appendix of $\mathscr{I}$. Thus, the off-diagonal density-matrix element corresponding to  the probe transition is obtained as
\begin{equation}
{\rho}_{a_1 c} = \tilde{\rho}_{a_1 c} \re^{- \ri \nu_p t} =
\frac{1}{Y \hbar} (\Omega_2^2 - 4 \Delta^2 +  2 \ri \gamma_2 \Delta)
\mathcal{E}_p \wp_{a_1 c} \re^{- \ri \nu_p t},
\label{eq:tilderhoa1c}
\end{equation}
where we have chosen, without loss of generality, $\theta_3=\pi/ 4$, 
$\theta_2=\pi/2+\pi/ 4$, moreover, $Y$ is defined to be
\begin{align}
Y = A + \ri B,
\end{align}
with
\begin{align}
A &= - 8 \Delta^3 +  2 \Delta (\Omega_1^2 \sin^2 \kappa x + \Omega_2^2 +
\Omega_3^2  )
\nonumber \\
&\qquad + 2 \gamma_1 \gamma_2 \Delta +
\Omega_1 \Omega_2 \Omega_3 (\re^{\ri \varphi} + \re^{-\ri \varphi}) \sin
\kappa x , \nonumber \\ B &= 4 \Delta^2 (\gamma_1 +\gamma_2) - 
\gamma_1 \Omega_2^2 - \gamma_2 \Omega_1^2 \sin^2 \kappa x.
\end{align}

The susceptibility at the probe frequency can be written as
\begin{equation}
\chi =\frac{2 N \wp_{a_1c} \rho_{a_1c} }{\epsilon_0 \mathcal{E}_p } \re^{\ri \nu_p t} 
      =\frac{2 N |\wp_{a_1c}|^2}{\epsilon_0 } \frac{ (\Omega_2^2 - 4 \Delta^2 + 2 \ri  \gamma_2 \Delta) }{Y \hbar},
\end{equation}
where $N$ is the atom number density in the medium. The real and imaginary
parts of susceptibility are given as
\begin{align}
\chi' & = \frac{2 N |\wp_{a_1 c}|^2 }{ \epsilon_0 \hbar Z} \{
(\Omega_2^2 - 4 \Delta^2) A + 2 \gamma_2 \Delta B\},\\ \chi'' & =
\frac{2 N |\wp_{a_1 c}|^2 }{ \epsilon_0 \hbar Z} \{ 2  \gamma_2
\Delta A - (\Omega_2^2 - 4 \Delta^2) B)\},
\label{Eq:Chiprime}
\end{align}
where $Z = Y Y^*$ and $\chi=\chi'+ \ri \chi''$. It is imperative to point out that the phase enters the susceptibility expression only through the quantities $A$ and $Y$. Even the phase dependence of $Y$ is only through the quantity $A$. Moreover, we observe that the phase 
dependent term in $A$ is $\Omega_1 \Omega_2 \Omega_3 (\re^{\ri \varphi} + \re^{-\ri \varphi}) \sin \kappa x $. Thus the phase factor could very well have come from either of the three driving fields. As pointed out earlier, if all the fields had phase dependence,
 only the collective phase would be important and no individual phase-dependent terms would occur. This is because  the Rabi frequencies $\Omega_{i}$ in all the other terms appear through $\Omega_{i}^{2}$, which is $|\Omega_{i}|^{2}$ for a complex Rabi 
frequency $\Omega_{i}=|\Omega_{i}| \re^{\ri \phi_{i}}$. The collective phase can be easily determined to be $\varphi=\phi_{2}+\phi_{3}-\phi_{1}$, by repeating the susceptibility calculation. Here $\phi_{i}$ is  the phase of the complex Rabi frequency $\Omega_{i}$ of the $i$th driving field.

In the next section we consider the imaginary part of the susceptibility
$\chi''$ in detail and obtain various conditions for subwavelength localization of the atom.

\section{Results and Discussions}
\label{Sec:Results}
We study the expression~(\ref{Eq:Chiprime}) for the imaginary part of the susceptibility on the probe transition in greater detail in the following 
discussion. It is clear that $\chi''$\mdash i.e., probe absorption\mdash depends 
on the controllable parameters of the system like probe field detuning,
amplitudes and phases of the driving fields. First we  present analytical considerations of the probe absorption maxima and its relation to the atom localization. Then we present the results of the numerical study for a variety of different sets of values of the parameters. In the end we present the dressed-state analysis to shed some light on the numerical results. 

\subsection{The probe absorption maxima}
Noting the dependence of $\chi''$ on $\sin{\kappa x}$, it is, in principle, possible to  obtain information
about the $x$ position of the atom as it passes through the cavity 
by measuring the probe absorption. Nevertheless, for precise localization of the
atom the susceptibility should show maxima or peaks at certain 
$x$ positions. We obtain the conditions for the presence of peaks in $\chi''$ in the discussion to follow. Eq.~(\ref{Eq:Chiprime}) can be rewritten as follows, 
using 
$\mathscr{N}={2 N |\wp_{a_1c}|^2}/{\hbar \epsilon_0}$, 
\begin{widetext}
\begin{align}
\frac{\chi''}{\mathscr{N}} &= 
\frac
   {A+ B(\kappa x)}
{
     \gamma_1\,[A + 2\, B(\kappa x)]
     + \gamma_2^2\,(\Omega_1^2\,\sin^2\kappa x -4\Delta^2)^2 
      + [8 \Delta^3 
- 2\,\Delta \,(\Omega_1^2\,\sin^2{\kappa x} + \Omega_2^2 + \Omega_3^2)
-  2\,\Omega_1 \Omega_2 \Omega_3\, \cos\varphi\, \sin{\kappa x}]^2
}
\nonumber \\
&=\frac
   {
   A+B(\kappa x)}{ \gamma_1\,[A + 2\, B(\kappa x) ]
   + \gamma_2^2 \Omega_1^4(\sin \kappa x - R_1)^2(\sin \kappa x - R_2)^2 + 4\Delta^2 \Omega_1^4(\sin \kappa x-R_3)^2(\sin\kappa x - R_4)^2}
\label{Eq:Rep2}
\end{align}
where
\begin{align}
A&= \gamma_1\,( 4\,\Delta^2\,\gamma_2^2 + 
  (\Omega_2^2-4\,\Delta^2)^2)\,, \nonumber \\
B(\kappa x)&=  \gamma_2
  ( \Omega_1^2\,\Omega_2^2\sin^2{\kappa x}\, + 
     4\,\Delta\, \Omega_1\,\Omega_2\,\Omega_3 \,\cos\varphi\,\sin{\kappa x}\,
     + 4\,\Delta^2\,\Omega_3^2)=\gamma_2 \, \Omega_1^2\,\Omega_2^2\,(\sin\kappa x-L_1)(\sin\kappa x - L_2)\,, \nonumber \\
   L_{1,2}&= \frac{2\,\Delta\,\Omega_3}{\Omega_1\,\Omega_2}
   {\left( - \cos\varphi  \pm 
      {\sqrt{\cos^2\varphi -1}} \right) }\,, \\ 
   R_{1,2}&=\mp\frac{2\,\Delta }{{{\Omega }_1}}\,,\nonumber \\
   R_{3,4}&=\frac{1}{2\,\Delta \,\Omega_1}\Biggl[-\Omega_2\,\Omega_3 \cos\varphi \pm
   \sqrt{\Omega_2^2\,\Omega_3^2 \cos^2\varphi - 4 \Delta^2(\Omega_2^2 + 
        \Omega_3^2- 4 \Delta^2)}\Biggr]\,.\nonumber
\end{align}
\end{widetext}
It can be seen that the probe field absorption would peak at positions satisfying 
\begin{equation}
\sin{\kappa x}=R_{1,2,3,4}\,.
\end{equation}
The roots $L_{1,2}$ do not contribute to the probe absorption maxima as they are the roots of the numerator as well, and these contributions mutually cancel.  Moreover, for a given set of parameters not all four roots contribute to the probe absorption maxima, as they have different weighting factors given by $\gamma_2 \Omega_1^4$ and $4\Delta^2\Omega_1$. The dominant weighting factor, being independent of position, governs which set of roots $\{R_{1,2}\}$ or $\{R_{3,4}\}$ will be important for dictating the atom-localization positions.

It can be clearly seen that for $\gamma_2=0$ the maxima positions gverned by $R_{1,2}$ do not occur; whereas, for $\Delta=0$ the roots $R_{1,2}$ are more important compared to $R_{3,4}$. In the regime where both $\Delta$ and $\gamma_2$ are zero interesting consequences follow. This competition of the roots gives rise to various interesting regimes of  parameters and possibilities in the atom localization. In the following we will throw light on the novel propreties arising due to this freedom. It can be noted that in $\mathscr{I}$ the roots $L_{1,2}$ and $R_{1,2}$ did not appear as $\gamma_2$ was taken to be zero, which leads to $B(\kappa x)=0$. For completeness we give the expression of $\chi''$ as used in $\mathscr{I}$:
\begin{widetext}
\begin{align}
\chi'' &= \frac{2 N |\wp_{a_1c}|^2}{\hbar \epsilon_0} 
\frac{\gamma_1 (\Omega_2^2 - 4 \Delta^2)^2}
{\gamma_1^2(\Omega_2^2 -4\Delta^2 )^2 + ( 8 \Delta^3 
- 2\,\Delta \,(\Omega_1^2\,\sin^2{\kappa x} + \Omega_2^2 + \Omega_3^2)
-  2\,\Omega_1 \Omega_2 \Omega_3\, \cos\varphi\, \sin{\kappa x})^2}
\nonumber \\
&=\frac{2 N |\wp_{a_1c}|^2}{\hbar \epsilon_0} 
\frac{\gamma_1 (\Omega_2^2 - 4 \Delta^2)^2}
{\gamma_1^2(\Omega_2^2-4\Delta^2 )^2 
+ 4 \Delta^2 \Omega_1^4\, (\sin{\kappa x}-R_3)^2(\sin{\kappa x}-R_4)^2}\,.
\label{Eq:chirootOld}
\end{align}
\end{widetext}
\forget{To illustrate the behavior of the probe absorption in detail for different parameter regimes, we performed extensive numerical study of the above expression; the results are summarized in the discussion to follow. To understand the numerical results we have also looked at the dressed-states approach  and provided qualitative explanations of the findings; they will be presented in the subsection to follow.}

A direct calculation of $\chi''$ from the equation~\eqref{Eq:Chiprime} shows that the positions of  maxima of  do not strongly depend on the decay parameters, and are function of only the drive field Rabi frequencies and phases. However, for a chosen value of the detuning the widths of the peaks observed in the plots of $\chi''$ vs $\kappa x$ depend on the values of the decay parameters.  To make connection with the positions of maxima predicted by the roots of the denominator in Eq.~\eqref{Eq:Rep2}, i.e, the roots $R_{1,2,3,4}$,  we study these roots in detail in the following discussion.

The probe field detunings required to obtain probe field absorption peaks as a function of the $x$ coordinate along the cavity field axis can be obtained by solving equations $\sin{\kappa x} = R_{1,2,3,4}$ for $\Delta$.   We denote the solutions  for $\sin \kappa x = R_{1,2}$  as $\delta_{1,2}$  and the solutions for $\sin \kappa x = R_{3, 4}$ as $\delta_{3,4,5}$.
It can be easily shown that
\begin{align}
\delta_{1,2} = \mp\frac{\Omega_1}{2}\sin\kappa x
\end{align}
and $\delta_{3,4,5}$ are the solutions of $\sin \kappa x = R_{3,4}$\mdash i.e., the cubic equation, 
\begin{equation}
4 \delta^3 - \delta(\Omega_1^2 \sin^2 \kappa x + \Omega_2^2 + \Omega_3^2) - \Omega_1 \Omega_2 \Omega_3 \sin \kappa x \cos \varphi = 0.
\label{Eq:delta345}
\end{equation}
When the relative phase $\varphi=\pi/2$, the above equation can be readily solved to give
\begin{align}
\delta_{3}&=0,\quad\delta_{4,5}= \pm \frac{1}{2}\, \sqrt{\Omega_1^2 \sin^2 \kappa x +  \Omega_2^2 + \Omega_3^2}\,.
\end{align}
Thus, for $\varphi=\pi/2$ the above  equations give the values of the probe detuning for observing probe absorption maxima as a function of the spatial position along the standing-wave field. It is clear that for $\delta=\Delta=0$ there is no  atom localization possible as the probe absorption would be the same at all spatial positions.  
We do not give expressions for $\delta_{3,4,5}$ for the case of $\varphi=0i$ as they are quite complicated; however, they can be readily evaluated numerically to verify the predictions.

It can also be noted that for the simplified case of $\Omega_2=\Omega_3=\Omega$ the $\delta_{3,4,5}$ expressions are considerably simplified and are given by
\begin{align}
\delta_{3}&= -\frac{1}{2}\,{\Omega_1 \sin \kappa x}\,, \nonumber \\ \delta_{4,5}&=\frac{1}{4}\,\[\Omega_1 \sin \kappa x \pm \sqrt{\Omega_1^2 \sin^2 \kappa x + 8 \Omega^2}\] \quad\text{ for } \varphi = 0.
\end{align}
\forget{and
\begin{align}
\delta_{3}&=0 \,, \nonumber \\
\delta_{4,5}&= \pm \frac{1}{2}\, \sqrt{\Omega_1^2 \sin^2 \kappa x + 2 \Omega^2}\quad\text{ for } \varphi = \pi/2.
\end{align}}
This means that $\delta_3=\delta_1$ for $\varphi = 0$;
however,  we also have 
\begin{align}
L_{1,2} &= -\frac{2 \Delta}{\Omega_1}\quad &\text{for } \varphi=0\,, \nonumber \\ 
L_{1,2} &= \frac{2 \Delta}{\Omega_1}(-1\pm \ri)\quad &\text{for } \varphi=\pi/2 \,.
\label{Eq:L12R}
\end{align}
Thus, $L_{1,2} = R_1$, therefore, the peaks arising from $\sin \kappa x = R_1$---i.e., $\delta_{1}$---will be completely suppressed for $\varphi=0$. Morever, $\delta_2$ will only appear if $\gamma_2$ is considerably larger compared to all other parameters of the system. Numerical study presented in the next subsection suggests that $\gamma_2>10\, \Delta_{\text{max}}$ for the roots $\delta_{1,2}$ to start showing up. It can also be seen that $\delta_{1}$ starts showing up for $\varphi=0$ if $\Omega_1 > 10\, \Omega_2$. These features can be understood by  observing  Eq.~\eqref{Eq:Rep2} and comparing the weighting coefficients of various roots of the numerator and denominator. This features are confirmed by the numerical study presented in the next subsection.

\forget{The results are as follows:
\begin{widetext}
\begin{align}
\sin \kappa x  &= R_{1,2} \Rightarrow \Delta=\pm \frac{\Omega_1}{2}\sin \kappa x= \delta_{1,2}(\kappa x) \nonumber \\
\sin \kappa x  &= R_{3,4} \Rightarrow \Delta=\delta_{3,4,5}(\kappa x) \nonumber \\
\text{where} \nonumber \\
\delta_{3} &= \frac{\sqrt[3]{3}\,({{{\Omega }_1}}^2\,{\sin \kappa x}^2 + {{{\Omega }_2}}^2 + 
 {{{\Omega }_3}}^2) + \sqrt[3]{D^2}
               }%
               {2\,\sqrt[3]{9D}}\nonumber \\
               
\delta_{4}&=\frac{}{4\,\sqrt[3]{9D}}
\text{with}\nonumber \\
D&={9\, \Omega_1\,\Omega_2\,
       \Omega_3\,\sin \kappa x\,\cos \varphi
       + {\sqrt{3}}\,
       {\sqrt{27\,\Omega_1^2\,\Omega_2^2\,
            \Omega_3^2\,{\sin^2 \kappa x}\,{\cos^2 \varphi}
             - {(\Omega_1^2\,\sin^2 \kappa x + 
               \Omega_2^2 + \Omega_3^2)}^3}}}
\end{align} 
\end{widetext}
}

\subsection{Numerical considerations}
In the discussion to follow we plots  the roots $\delta_{1,2,3,4,5}$ as a function of $\kappa x$ and show their connection with the behavior of $\chi''$ vs the probe detuning along the cavity field.

\begin{figure*}
\centerline{\includegraphics[width=0.96\textwidth]{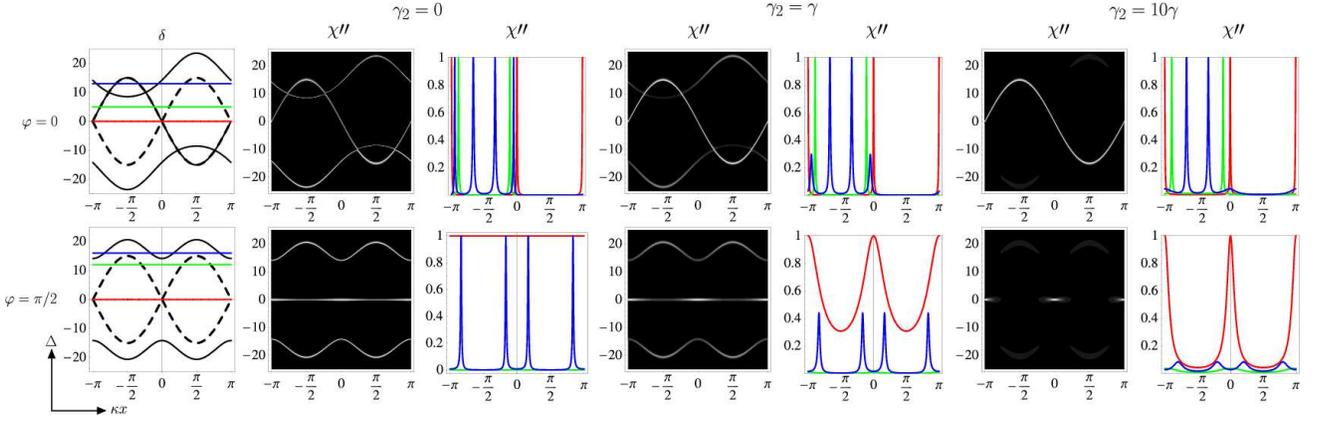}}
\caption{\label{Fig:effectofgamma} The effect of $\gamma_2$ on the localization.
The parameters are $\Omega_1=30, \Omega_2=\Omega_3=20$ $\gamma_1=1$. Top row: $\varphi=0$ and Bottom row: $\varphi=\pi/2$.  The first column shows the plots of the roots $\delta_{1,2}$ using dashed lines and that of $\delta_{3,4,5}$ by solid lines. The colored horizontal lines correspond to values of detuning $\Delta$ chosen to plot $\chi''$ vs $\kappa x$ in the line-plots shown later in each row for different values of $\gamma_2$ shown on top. The line plots are preceded by contour plots of $\chi''$ to give an idea of its dependence on $\Delta$ as well as position, $\kappa x$. The brightness of a given location in the contour plot is proportional to its height in a 3D plot of $\chi''$ vs $\Delta$ and $\kappa x$.  It can observed that for $\Omega_2=\Omega_3=\Omega$ the roots $\delta_1$ and $\delta_3$ coincide. The roots $\delta_{4,5}$ loose their significance as $\gamma_2$ increases. This result can be explained through the dressed states approach as described in the text. For $\Delta=0$ and $\gamma_2=10\gamma$ the root $\delta_{1}=\delta_{3}$ dominates as opposed to $\delta_{4,5}$ as expected. Despite expectation the root $\delta_2$ never appears for $\varphi$ different from $\pi/2$. It can be noticed that for $\Delta=0$ and non-zero $\gamma_2$ one might expect both the roots $\delta_{1,2}$ to dominate, however, this is case only when $\varphi=\pi/2$.}
\end{figure*}

To make contact with our earlier work, $\mathscr{I}$,  we first consider the parameter range with $\gamma_2=0$ and $\Omega_2=\Omega_3$ and study the effect of increasing $\gamma_2$ on that result. The findings are summarized in 
Fig.~\ref{Fig:effectofgamma}. In the first column we plot the roots $\delta_{1,2,3,4,5}$ so that their relation to the probe absorption maxima can be established. Then the contour-density plot of $\chi''$ vs the probe-field detuning $\Delta$ and the $\kappa x$ and $\chi''$ vs $\kappa x$ for chosen value of $\Delta$ are plotted for different values of  $\gamma_2$ starting with $\gamma_2=0$. The color of the line plots corresponds to the horizontal lines in first-column plots for the respective phase value. This correspondence helps to determine the positions and number of the peaks in the line plots from the places at which the horizontal line intersects the roots $\delta_{1,2,3,4,5}(\kappa x)$. It is clear that the roots $\delta_{3,4,5}$, denoted by solid lines, are dominant most of the times as opposed to $\delta_{1,2}$. The same conclusion can be drawn from the dressed-states approach but for different reasons as discussed in the next subsection.

We label the dominant roots, $\delta_{3,4,5}$, such that the root crossing the $\Delta=0$  line (for $\varphi=0$) or the $\Delta=0$ line itself  (for $\varphi=\pi/2$) as $\delta_{3}$, the root above the $\Delta=0$ line  as $\delta_{4}$, and the one below the $\Delta=0$ line as $\delta_{5}$ as seen in the first-column plots in Fig.~\ref{Fig:effectofgamma}. Among the non-dominant roots---denoted by dashed lines in the first-column plots in Fig.~\ref{Fig:effectofgamma}---the root that coincides with $\delta_{3}$ is denoted as $\delta_1$ and the other one is $\delta_2$. It can be noted that the $\delta_{1,2}$ are independent of the relative phase of the drive fields $\varphi$.  This labeling of the roots will be used for the rest of the discussion. 

In Fig.~\ref{Fig:effectofgamma}, we further observe that with increasing $\gamma_2$, the roots $\delta_{4,5}$ start diminishing. It is, however, to be noted that this behavior can only be seen from the density plots of $\chi''$ and the plots of $\delta_{3,4,5}$ themselves do not give this information. Another way to explain the peak widths and their dominance is through the decay rates of the dressed-states. We discuss the implications in the next subsection where we evaluate the dressed-states.  

We consider the results depicted in the line-plots in Fig.~\ref{Fig:effectofgamma} in further detail. The effect of increasing $\gamma_2$ can be easily seen from the peaks arising due to the roots $\delta_{4,5}$, as seen in the blue plots in Fig.~\ref{Fig:effectofgamma} for both the values of $\varphi=0,\pi/2$.  We first consider the results for $\varphi=0$---For $\gamma_2=0$, out of the four blue peaks (corresponding to $\Delta=13\gamma$) occurring the first half-wavelength region, the outer ones arise from $\delta_4$ and the inner ones from $\delta_{3}$. Thus, the expectation--from the density plots--would be that the inner roots would remain sharp and dominant while the outer ones will loose their height and sharpness; this expectation is confirmed by the line-plots for $\gamma_2=\gamma$ and $\gamma_2=10\gamma$. The green ($\Delta=5\gamma$) peaks which arise solely through $\delta_{3}$ are unaffected by increasing $\gamma_2$. The same is true for the red line-plots  which correspond to the probe detuning of $\Delta=0$ giving rise to peaks at the nodes of the cavity standing-wave field. Now we consider the case of $\varphi=\pi/2$---Here $\delta_{3}$ coincides with the zero line, hence for $\gamma_2=0$ the red line-plot ($\Delta=0$) gives equal absorption at all spacial points but starts showing spacial dependence as $\gamma_2$ increases which can also be clearly seen from the density plots in the $\Delta=0$ region. The blue ($\Delta=16\gamma$) peaks, in this case, arise from $\delta_{4}$ and therefore diminish in height as $\gamma_2$ increases.  The green ($\Delta=12\gamma$) plots show that $\delta_{1,2}$ do not contribute the this particular choice of parameters and show zero absorption for all values of $\gamma_2$. The results depicted in the line plots coincide very well with the corresponding density plots.

\begin{figure*}
\centerline{\includegraphics[width=0.96\textwidth]{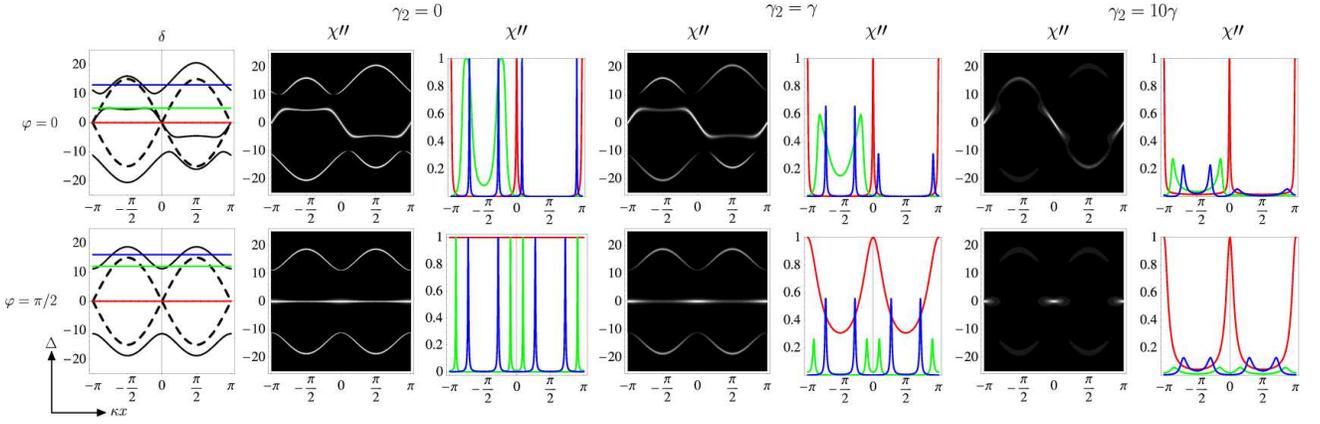}}
\caption{\label{Fig:interplay} Localization characteristics for non-identical drive field intensities and interplay of different roots. The parameters are  $\Omega_1=30, \Omega_2=20, \Omega_3=10$ $\gamma_1=1$. Top row: $\varphi=0$ and bottom row: $\varphi=\pi/2$. The structure of the figure is the same as Fig.~\ref{Fig:effectofgamma}. The roots $\delta_{4,5}$ loose their significance as $\gamma_2$ increases. This result can be explained through the dressed states approach as described in the text. For $\Delta=0$ and $\gamma_2=10\gamma$ the roots $\delta_{1,2}$ dominate as opposed to $\delta_{3,4,5}$ as expected for $\varphi=\pi/2$. This can be clearly seen  as $\delta_3=0$ line becomes insignificant and only the nodal points (arising from $\delta_{1,2}$) remain with increasing $\gamma_2$.}
\end{figure*}

The red curves in Fig.~\ref{Fig:effectofgamma}, corresponding to the detuning $\Delta=0$, show different behavior for different  phase values. For $\varphi=0$, the height and width of the peaks observed at the zero detuning of the probe field are immaterial of the lifetime of level $\ket{2}$, as they arise from the zero eigenvalue of the dressed state which does not contain any $\ket{2}$ component. This corresponds to a regime of localization that is very common in several other localization proposals, namely, observance of localization peaks at the nodes of the standing-wave cavity field.  For $\varphi=\pi/2$, the $\Delta=0$ value is special as it does no show any localization for $\gamma_2=0$ (observe the red plots in the  lower row of the Fig.~\ref{Fig:effectofgamma}). It can be seen that these peaks become sharper with increasing $\gamma_2$, whereas the green and blue plots still show dimishing height and increasing width of the peaks. This can be explained as follows: at $\Delta=0$ and $\gamma_2\neq0$ roots $\delta_{1,2}$ dominate as opposed to $\delta_{3,4,5}$ for all other values of the detunings. Thus, for $\varphi=\pi/2$, the root $\delta=0$ starts losing its significance as   $\gamma_2$ increases and only the nodal points show peaks which arise from $\delta_{1,2}$ shown by dashed-line plots in first column. This feature is absent for $\varphi=0$ as $\delta_{1}$ does not occur for the parameters of  Fig.~\ref{Fig:effectofgamma}---as explained in the context of Eq.~\eqref{Eq:L12R}---and $\gamma_2$ is not large enough for $\delta_{2}$ to show up compared to the dominant root $\delta_{3}$.  However, with a different parameter range of values this competition of roots can be seen for $\varphi=0$ as depicted in Fig.~\ref{Fig:interplay}. It is also interesting to note that for $\varphi=0$, in the current figure, the localization peaks at the nodes of the cavity standing-wave field are much sharper than the ones observed for $\varphi=\pi/2$. 

Thus, the general conclusion that can be drawn from Fig.~\ref{Fig:effectofgamma} is  that as $\gamma_2$ increases only the maxima due to the root $\delta_3$ (for $\varphi=0, \pi$) and root $\delta_{1,2}$  (for $\varphi=\pi/2$)  show sharp peaks and the other peaks diminish in magnitude and sharpness. The drive field parameters chosen in Fig.~\ref{Fig:effectofgamma} were as that of the earlier work in $\mathscr{I}$ except for the non-zero $\gamma_2$. 

Now we study the effect of varying the amplitudes of the drive fields and go beyond the condition $\Omega_2=\Omega_3$ on the probe field absorption. The results are summarized in Fig.~\ref{Fig:interplay}.
The results for $\varphi=\pi/2$ in Fig.~\ref{Fig:interplay} are very similar to the one in  Fig.~\ref{Fig:effectofgamma}, except for the green ($\Delta=12\gamma$) plots. Here the detuning values are the same in both the figures, however, the roots $\delta_{4,5}$ have a larger range than before due to their dependence on the drive-field Rabi frequencies. Thus, new roots appear for the green plots in Fig.~\ref{Fig:interplay} as opposed to  no roots in Fig.~\ref{Fig:effectofgamma}. These roots however diminish as $\gamma_2$ is increased making them useless for atom localization for larger $\gamma_2$. For the case of $\varphi=0$ the roots $\delta_{3,4,5}$ have completely different profiles compared to their counterparts in Fig.~\ref{Fig:effectofgamma}.
The red plots have the same behavior as in Fig.~\ref{Fig:effectofgamma} being a very commonly observed localization regime for the probe detuning $\Delta=0$.   Moreover, the disappearance of the roots $\delta_{4,5}$ with increasing $\gamma_2$ exists in this parameter range as well. Due to this the blue and green peaks loose their height and sharpness with increasing $\gamma_2$. 

We observe that an interesting regime arises when $\gamma_2\neq 0$ and the detuning of the probe field  $\Delta=0$. In this regime the roots $\delta_{1,2}$ dominate compared to $\delta{3}=0$ for $\varphi=\pi/2$; it can be clearly seen from the last column of plots in Fig.~\ref{Fig:interplay}. The significance of this is clearly apparent for $\varphi=\pi/2$, where the probe absorption is uniform over all spatial points for $\delta_{3}=0$ when $\gamma_2=0$; however as $\delta_{1,2}$ become dominant due to increasing $\gamma_2$ absorption peaks start emerging at positions corresponding to the nodes of the standing-wave field.  The behavior for $\varphi=0$ is a bit different and it needs complete dressed-states analysis to explain that has been found to be  quite complicated.

Another interesting feature observable from Fig.~\ref{Fig:interplay} is that the root given by $\delta_{3}$ is not dominant at all spacial positions as it is in Fig.~\ref{Fig:effectofgamma}. With increasing $\gamma_2$ the relatively flat regions in the plot of $\delta_{3}$ vs $\kappa x$ start loosing their significance as $\gamma_2$ is increased. This can be ascribed to the broadening of the resonances owing to increased $\gamma_2$. When $\delta_{3}$ remains close close to the line $(\Omega_1/2) \sin \kappa x$, the state is very close to the first eigenstate discussed in Eq.~\ref{Eq:eigenphi0} and these parts remain sharp $\gamma_2$ does not affect the sharpness. Whereas, departure of $\delta_{3}$ from $(\Omega_1/2) \sin \kappa x$ lines can be ascribed to increasing components of state $\ket{a_2}$, which decays with $\gamma_2$, in the dressed state. Thus, the flat regions loose their significance for localization with increasing $\gamma_2$.

\begin{figure}[ht]
\centerline{\includegraphics[width=0.95\columnwidth]{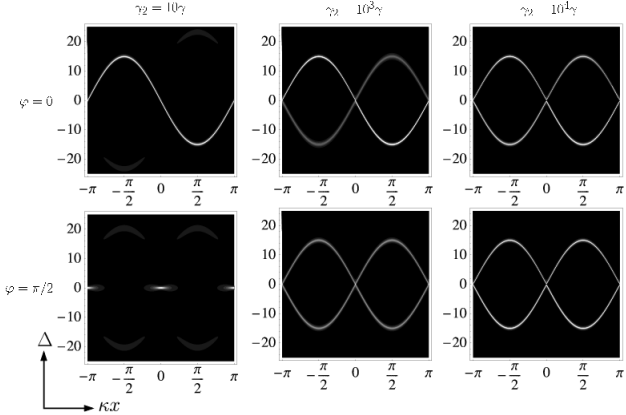}}
\caption{\label{Fig:Unrealistic} Dominance of the roots $R_{1,2}$. The parameter values are same as in Fig.~\ref{Fig:effectofgamma} except for $\gamma_2$. $(a)$ $\gamma_2=10\gamma$. $(b)$ $\gamma_2= 10^3\gamma$ $(c)$ $\gamma_2=10^4\gamma$. It can be seen that the less dominant maxima slowly vanish as $\gamma_2$ is increased.  Thus for large $\gamma_2$ compared to $\gamma_1$ only $\Delta=0$ shows probe peak absorption. As $\gamma_2$ increases $R_1$ starts dominating which is the same as $R_3$, however for somewhat unrealistically larger $\gamma_2$, $R_{3,4}$ roots completely dominate. The apparance of $R_4$ can not be explained by the dressed state approach.} 
\end{figure}

Noting that for $\gamma_2\gg\Delta$  we can expect the behavior of the probe absorption to be completely dominated by $\delta_{1,2}$ as opposed to $\delta_{3,4,5}$ we choose appropriate values for the parameters and consider the density plots of $\chi''$ in Fig.~\ref{Fig:Unrealistic}. 
However, this parameter range is unrealistic and also not very useful as there will be four peaks observed for the detuning lying in the interesting regime as both the $\delta_{1,2}$
roots exist giving rise to four intersection points in one wavelength for a chosen value of the probe detuning. Nevertheless, the observations made in the context of Eq.~\eqref{Eq:L12R} can be confirmed from the results in Fig.~\ref{Fig:Unrealistic}. 
Both the curves appearing for $\varphi=\pi/2$ have equal characteristics as they both are due to $\delta_{1,2}$, which arise from the same term in the denominator. However, for $\varphi=0$, as discussed earlier, $\delta_{1}$ is cancelled by the roots of the numerator and instead of $\delta_{1}$, the sharper root $\delta_{3}$ appears. Only when $\gamma_2$ is sufficiently large $\delta_{3}$ and $\delta_{2}$ acquire same sharpness as seen in the last column plots of Fig.~\ref{Fig:Unrealistic}.

Deeper understanding of the interplay of different roots can be achieved through the dressed states calculation. We determine the dressed states in the next subsection and explain the above obtained results from a different point of view.

\subsection{Dressed-states approach}
To understand the emergence of several roots for the maxima of the probe absorption we consider the dressed states approach. The effective Hamiltonian, taking into account only the strong drive fields, can be expressed as
\begin{widetext}
\begin{equation}
\mathscr{H}_{\rm eff}=
\frac{\ri}{2} \(
\begin{matrix}
0 & \Omega_3\,\re^{-\ri \varphi}\,\re^{\ri k x \cos \theta_3}& \Omega_1\,\sin \kappa x\\
\Omega_3\,\re^{\ri \varphi}\,\re^{-\ri k x \cos \theta_3} & 0 & \Omega_2\,\re^{\ri k x \cos \theta_2}\\
 \Omega_1\,\sin \kappa x & \Omega_2\,\re^{-\ri k x \cos \theta_2}  & 0
\end{matrix}
\)
\label{Eq:EffHam}
\end{equation}
\end{widetext}
in the basis
 $\{ \ket{a_1}, \ket{a_2}, \ket{b}\} $.
Choosing $\theta_2=\pi/4$ and  $\theta_3=\pi/2 + \pi/4$, in the above Eq.~\eqref{Eq:EffHam} we arrive at the secular equation
\begin{equation}
4 \lambda^3 - \lambda(\Omega_1^2 \sin^2 \kappa x + \Omega_2^2 + \Omega_3^2) - \Omega_1 \Omega_2 \Omega_3 \sin \kappa x \cos \varphi = 0,
\label{Eq:secular}
\end{equation}
where $\lambda$ are the eigenenergies of the Hamiltonian. 
It can be noted that Eq.~\eqref{Eq:secular} is identical to Eq.~\eqref{Eq:delta345}. Thus, there is a direct connection between the detuning values for  the probe field at which it experiences maximum absorption, $\delta_{3,4,5}$, and the dressed state eigenvalues. The dressed state eigenvalues $\lambda$ give the Stark-shifts in the energy of the state $\ket{a}$. When this Stark-shifted transition $\ket{a}$--$\ket{c}$ is probed by the weak probe field, the resonances will occur at the points where the probe frequency matches the energy level difference between the Stark-shifted levels $\ket{a}$ and level $\ket{c}$.  If the probe field frequency, or the detuning $\Delta$, is chosen such that it is in resonance with one of the dressed states then it experiences absorption maxima. This can be expressed by a condition $\lambda=\delta$. It can, however, be noted that only the detuning solutions $\delta_{3,4,5}$ can be explained through the dressed-states approach and not the solutions $\delta_{1,2}$, as explained later.

Actual form of the dressed states for general parameters is quite complicated and in not required as we are only interested in locating the positions of the resonances in the frequency space.  In general the probe absorption peaks are quite sharp except at the  stationary points along the $x$ axis and when $\gamma_2$ increases. This calculation can be extended further to obtain the spontaneous decay rates of the dressed states. These decay rates could then give more information about the widths of the probe absorption maxima and the loss of sharpness with increasing $\gamma_2$.

For a completely general set of parameters evaluating the actual form of the dressed states and their spontaneous decay rates is sufficiently involved compared to the information that can be gained by such an exercise.   Therefore, we consider a restricted regime of parameters to extract information about  the sharpness of the probe absorption peaks through the dressed states approach. 
Assuming $\Omega_2=\Omega_3=\Omega$  and $\varphi=0$ we  obtain the eigenvalues to be
\begin{align}
\left\{ -\frac{1}{2}{\Omega_1\,\sin\kappa x }\,, 
\frac{1}{4}\({\Omega_1\sin\kappa x\, \pm 
    {\sqrt{8\,{\Omega }^2 + \Omega_1^2
        \sin^2 \kappa x}}}\)\right\}
\end{align}
with the corresponding eigenstates
\begin{align}
\label{Eq:eigenphi0}
\frac{1}{\sqrt{2}}\(
\begin{matrix}
-\ket{a_1} \\
0 \\
\ket{b}
\end{matrix}
\),\quad
\mathscr{N}^{(\pm,0)}\(
\begin{matrix}
\ket{a_1} \\
c_{2}^{(\pm,0)}\ket{a_2}\\
\ket{b}
\end{matrix}
\)
\end{align}
where
\begin{equation}
c_{2}^{(\pm,0)}=
\frac{\re^{\frac{-\ri \,k\,x}{{\sqrt{2}}}}\Omega \,\left( 3\Omega_1\sin\kappa x\,
        \pm
      {\sqrt{8\,{\Omega }^2 +  \Omega_1^2
        \sin^2 \kappa x}} \right) }
    {2\,{\Omega }^2 +  \Omega_1^2
        \sin^2 \kappa x \pm
      \Omega_1\sin\kappa x\,
       {\sqrt{8\,{\Omega }^2 +  \Omega_1^2
        \sin^2 \kappa x}}}\,
\end{equation}
and $\mathscr{N}^{(\pm,0)}$ is the appropriate normalization constant.
Whereas, for $\varphi=\pi/2$ we obtain the eigenvalues 
\begin{align}
\{0, \pm\frac{1}{2}\sqrt{2 \Omega^2 + \Omega_1^2 \sin\kappa x}\}
\end{align}
with the corresponding eigenstates
\begin{align}
\mathscr{N}_1^{(\pi/2)}\(
\begin{matrix}
\ri \ket{a_1} \\
-\ri\frac{1}{\Omega} \re^{\frac{-\ri \,k\,x}{\sqrt{2}}}\Omega_1 \sin\kappa x \ket{a_2} \\
\ket{b}
\end{matrix}
\),\nonumber \\
\mathscr{N}_2^{(\pm,\pi/2)}\(
\begin{matrix}
c_1^{(\pm, \pi/2)}\ket{a_1} \\
c_2^{(\pm, \pi/2)}\ket{a_2} \\
\ket{b}
\end{matrix}
\)
\end{align}
where
\begin{align}
c_1^{(\pm, \pi/2)} = \frac{-\Omega_1 \sin \kappa x \pm \ri \sqrt{2 \Omega^2 + \Omega_1^2 \sin^2 \kappa x}}{\ri\, \Omega_1 \sin \kappa x \mp \sqrt{2 \Omega^2 + \Omega_1^2 \sin^2 \kappa x}} \nonumber \\
c_2^{(\pm, \pi/2)} =\frac{- 2\, \Omega\, \re^{\frac{-\ri \,k\,x}{\sqrt{2}}}}{\ri\, \Omega_1 \sin \kappa x \mp \sqrt{2\, \Omega^2 + \Omega_1^2 \sin^2 \kappa x}}
\end{align}
with $\mathscr{N}_1^{(\pi/2)}$ and $\mathscr{N}_2^{(\pm,\pi/2)}$ being the appropriate normalization constants.

The message to be taken from the dressed states representation $\mathscr{N} (c_{a_1} \ket{a_1} + c_{a_2} \ket{a_2} + c_b \ket{b})$  in the bare atomic levels  is that the decay rate of the corresponding dressed-state is given by $\gamma= |c_{a_1}|^2 \gamma_1 + |c_{a_2}|^2 \gamma_2$, as the level $\ket{b}$ is the ground state. Therefore, it is clear the for $\varphi=0$ the first dressed-state has the deay rate $\gamma_1/4$ whereas the other states decay at the rate  proportional to $|\mathscr{N}^{(\pm,0)}|^2 (\gamma_1 + |c_{2}^{(\pm,0)}|^2 \gamma_2)$. Resulting in sharp localization peaks when the probe field is in resonance with the first dressed state and not so sharp localization peaks when the probe field is in resonance with the other two dressed states.  In fact with increasing $\gamma_2$, as seen already in the numerical solutions, the latter two states contribute wider and wider resonances which are increasingly useless for atom localization.  Similar conclusions can be drawn for the case of $\varphi=\pi/2$; all three roots are equally sharp when $\gamma_2=0$ and the latter two roots increasingly loose their sharpness and decrease in amplitude for larger $\gamma_2$. This observation can be confirmed through the plots in Figs.~\ref{Fig:effectofgamma} and~\ref{Fig:interplay}.

\begin{figure}
\centerline{\includegraphics[width=0.95\columnwidth]{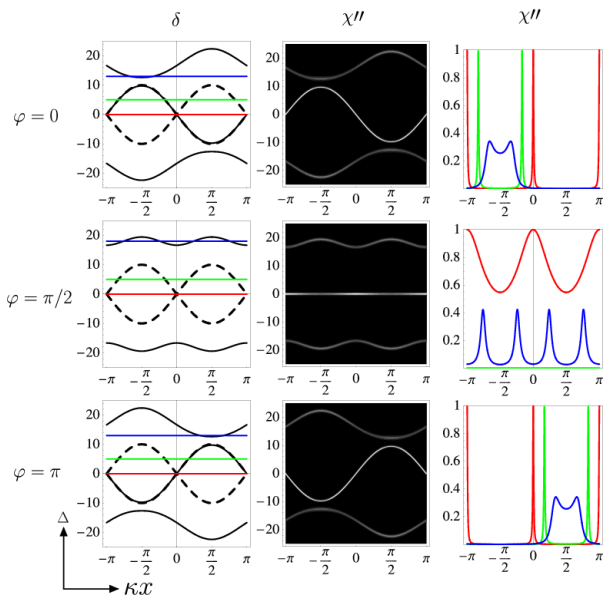}}
\caption{\label{Fig:GSHWL}  Illustrating the appropriate conditions to obtain good sub-half-wavelength probe absorption peaks along the cavity field\mdash i.e., sub-half-wavelength localization (observe the green plots in last column.). The parameters are $\Omega_1=20\gamma$, $\Omega_2=22\gamma$, $\Omega_3=25\gamma$. $\gamma_1=\gamma_2=\gamma$. In the first column we show the plots of the roots $\delta_{1,2}$ in dashed lines and that of the roots $\delta_{3,4,5}$ in solid lines alongwith horizontal lines for chosen values of probe detuning $\Delta$ that will be considered for more study in column 3.
In the second column we show the density plots of $\chi''$ and in the third column we choose $\Delta=5\gamma$ (green) and show the probe absorption peaks to illustrate the regime of sub-half-wavlength localization and its dependence on the phase $\varphi=0$. We choose other $\Delta$ values as well shown in blue and red to contrast the sub-half-wavelength regime. Note $\Omega_2, \Omega_3 > \Omega_1$ seperates the central root, $\delta_{3}$, from the other ones $\delta_{4,5}$ and provides a phase dependent localization for $\varphi=0$ and no localization for $\varphi=\pi/2$. The localization peaks would appear in the second sub-half-wavelength region if $\varphi=\pi$.}
\end{figure}

Another important conclusion that can be drawn from the dressed-state eigenvalues is that for the case of $\varphi=0$ the eigenvalues can be made to be well separated by choosing $\Omega_1$ to be little smaller than $\Omega_2=\Omega_3=\Omega$. In such a case the three roots do not overlap and the detuning can be chosen in the range $\{0, \Omega_1/2\}$ to obtain sub-half-wavelength localization. We illustrate this regime in Fig.~\ref{Fig:GSHWL}.

Moreover, this result holds true even when the drive fields $\Omega_2$ and $\Omega_3$ do not have the same value, $\Omega_2\neq\Omega_3$, and when $\Omega_1<\Omega_{2},\Omega_3$. Another message to be taken from Fig.~\ref{Fig:GSHWL} is that the results for $\varphi=\pi$ are mirror image of that of $\varphi=0$ taken around the vertical line $\kappa x = 0$. This holds true for all parameter ranges, hence we have plotted only the non-trivial cases $\varphi=0$ and $\varphi=\pi/2$ in all the other plots. The range of detuning $\Delta$ spanned by the root $\delta_{3}$ (for $\varphi=0,\pi$) gives an ideal range where sub-half-wavelength localization can be observed, which can always be calculated by solving the Eqs.~\eqref{Eq:secular} or~\eqref{Eq:delta345}, when it is very well separated from the other roots $\delta_{4,5}$. This happens as discussed above when $\Omega_1 < \Omega_2,\Omega_3$.

In spite of the use of the restricted parameters  for the evaluation of the dressed-states the results are valid in general as our numerical studies show. Neverthelsess, it can be noted that the parameter range where $\gamma_2$ has a role to play on the dominance of the roots $\delta_{1,2}$ or when $\gamma_2\gg\Omega_{1,2,3}$ can not be explained through the dressed-states approach (See Fig.~\ref{Fig:Unrealistic}). This breakdown of the dressed-states approach for large $\gamma_2$  is easy to understand. Dressed-state calculation is usually done with the assumption that  the Drive field Rabi frequencies are large compared to all the other parameters of the system, which breaks down in the large $\gamma_2\gg\Omega$ limit, giving rise to roots which are not predictable by the dressed-states. 

\section{Conclusions}
We have studied of  a variant of a $\Lambda$-type EIT, where a phase dependence is introduced through the application of three driving fields in a loop-configuration. The advantage of the phase dependence is in the tunability that becomes available to manipulate the response of the atomic medium to a weak probe field. By choosing one of the drive fields to be a standing-wave field of the cavity the phase dependence can be extended to obtain atom localization. We have given equations that could be used to simulate several, apparently quite different, energy level schemes. Effect of different parameters are studied with analytical as well as numerical techniques. A dressed-states approach is developed and it is used to explain the peak probe absorption and the peak widths. Also a region of parameters is identified which gives clean sub-half-wavelength localization for a wide range of probe detunings; thus, increasing the applicability of the model. In this range of parameters we show how the choice of phase governs whether localization would be observed or not.

\acknowledgments
Part of this work was carried out (by K.T.K.)
at the Jet Propulsion Laboratory under 
a contract with the National Aeronautics and Space Administration (NASA). 
K.T.K. acknowledges support from the National Research Council and
NASA, Codes Y and S.  M.S.Z. acknowledges  support of the Air
Force Office of Scientific Research,
DARPA-QuIST, TAMU
Telecommunication and Informatics Task Force (TITF) Initiative,
and the Office of
Naval Research. 

\forget{\bibliographystyle{prsty}
\bibliography{AtomLocalization}}

\forget{
}

\end{document}